\documentclass[11pt,letterpaper]{article}
\pdfoutput=1
\usepackage{braket}
\usepackage{jheppub}
\usepackage{multirow}
\usepackage{subcaption}
\usepackage{tikz}
\usetikzlibrary{arrows.meta}

\newcommand{\e}{\epsilon}

\title{Landau Singularities of the 7-Point Ziggurat I}

\author[a]{Luke Lippstreu,}
\author[a,b]{Marcus Spradlin,}
\author[a]{and Anastasia Volovich}

\affiliation[a]{Department of Physics, Brown University,\\
	182 Hope Street, Providence, RI 02912, U.S.A.}
\affiliation[b]{Brown Theoretical Physics Center, Brown University,\\
	340 Brook Street, Providence, RI 02912, U.S.A.}
 
\emailAdd{luke\_lippstreu@brown.edu}
\emailAdd{marcus\_spradlin@brown.edu}
\emailAdd{anastasia\_volovich@brown.edu}

\abstract{We compute the leading (first-type Landau) singularities of a certain four-loop 7-point graph that is related to the 7-point ``ziggurat'' graph by the graphical moves familiar from equivalent circuit theory. We find perfect agreement with a subset of the ``heptagon symbol alphabet'' that has appeared in the context of planar $\mathcal{N}=4$ super-Yang-Mills theory. The remaining heptagon symbol letters are found in its subleading Landau singularities, which we address in a companion paper.}

\begin{document} 
\maketitle

\section{Introduction}

A key aspect of the S-matrix program is the expectation that scattering amplitudes should be largely determinable from a thorough understanding of their analytic structure. In recent years planar $\mathcal{N}=4$ super-Yang-Mills (SYM) theory has served as an exemplar of this approach~\cite{Arkani-Hamed:2022rwr}. In particular, the assumption that the singularities of all six- and seven-point amplitudes are encoded in symbol letters that are cluster variables~\cite{Goncharov:2010jf,Golden:2013xva}, together with physical input from near-collinear and multi-Regge kinematics, has allowed them to be determined to high loop order (see~\cite{Caron-Huot:2020bkp} for a review).

While it is known that higher-point amplitudes in SYM theory (and certainly those in other, less ``simple'' field theories) can have significantly more complicated analytic structure, a general criterion for determining the locations of singularities of Feynman integrals was formulated over 60 years ago by Landau~\cite{Landau:1959fi}; see Sec.~1 of~\cite{Fevola:2023fzn} for a thorough historical overview. In recent years it has proven fruitful to explore general implications of the Landau equations and other classic work on discontinuities of amplitudes in the context of the modern amplitudes program; see for example~\cite{Dennen:2015bet,Dennen:2016mdk,Prlina:2017azl,Prlina:2017tvx,Bourjaily:2020wvq,Mizera:2021fap,Mizera:2021icv,Hannesdottir:2021kpd,Hannesdottir:2022xki, Dlapa:2023cvx, Gardi:2022khw}. High loop order and certain all-order analyses of the Landau equations for particular graphs and amplitudes can be found in~\cite{Kolkunov1,Kolkunov2,Stapp,Correia:2021etg}. Important recent progress in using computer algebra systems to solve Landau equations has been discussed in~\cite{Mizera:2021icv, Fevola:2023fzn, Fevola:2023kaw}.

Knowing the locations of an amplitude's singularities is closely related to, but not quite the same as, knowing its symbol alphabet. Knowledge of the former only provides information about where symbol letters \emph{vanish} (or where the letters themselves have algebraic branch points, as a function of the kinematic data), not necessarily what the symbol letters \emph{are} (although in many cases it gives enough information to make a natural guess).

\begin{figure}
\centering
\begin{subfigure}[t]{0.2\textwidth}
\centering
\includegraphics{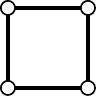}
\caption{}
\end{subfigure}
\begin{subfigure}[t]{0.2\textwidth}
\centering
\includegraphics{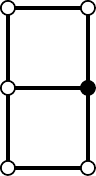}
\caption{}
\end{subfigure}
\begin{subfigure}[t]{0.28\textwidth}
\centering
\includegraphics{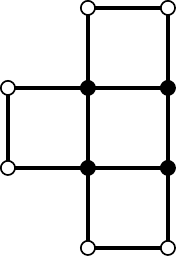}
\caption{}
\end{subfigure}
\begin{subfigure}[t]{0.28\textwidth}
\centering
\includegraphics{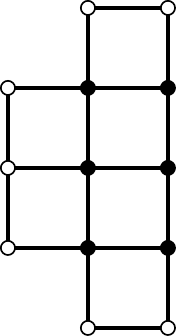}
\caption{}
\end{subfigure}
\caption{The four-, five-, six- and seven-point ziggurat graphs. A massless external leg attaches to each white vertex. The pattern continues by adding another column to the right each time $n$ increases by 2.}
\label{fig:one}
\end{figure}

\begin{figure}
\begin{subfigure}[b]{0.45\textwidth}
\centering
\begin{tikzpicture}[scale=0.75]
\draw[line width=0.6mm] (0,0) circle (2cm);
\draw[line width=0.6mm] (0,2) -- (0,0);
\draw[line width=0.6mm] (0,0) -- (+1.73205,-1);
\draw[line width=0.6mm] (0,0) -- (-1.73205,-1);
\draw[line width=0.6mm] (1.96962, 0.347296) -- (2.95442, 0.520945);
\draw[line width=0.6mm] (1.28558, 1.53209) -- (1.92836, 2.29813);
\draw[line width=0.6mm] (-1.28558, 1.53209) -- (-1.92836, 2.29813);
\draw[line width=0.6mm] (-1.96962, 0.347296) -- (-2.95442, 0.520945);
\draw[line width=0.6mm] (0.517638, -1.93185) -- (0.776457, -2.89778);
\draw[line width=0.6mm] (-0.517638, -1.93185) -- (-0.776457, -2.89778);
\node (10) at (0,3.5) {\color{white}$7$\color{black}};
\node (7) at (-1.51859, -3.15339) {\color{white}$4$\color{black}};
\end{tikzpicture}
\caption{${\cal{G}}_6$}
\end{subfigure}
\begin{subfigure}[b]{0.45\textwidth}
\centering
\begin{tikzpicture}[scale=0.75]
\draw[line width=0.6mm] (0,0) circle (2cm);
\draw[line width=0.6mm] (1.56366, 1.24698) -- (2.34549, 1.87047);
\draw[line width=0.6mm] (1.94986, -0.445042) -- (2.92478, -0.667563);
\draw[line width=0.6mm] (0.867767, -1.80194) -- (1.30165, -2.70291);
\draw[line width=0.6mm] (-0.867767, -1.80194) -- (-1.30165, -2.70291);
\draw[line width=0.6mm] (-1.94986, -0.445042) -- (-2.92478, -0.667563);
\draw[line width=0.6mm] (-1.56366, 1.24698) -- (-2.34549, 1.87047);
\draw[line width=0.6mm] (0,-2) -- (0,3);
\draw[line width=0.6mm] (-2,0) -- (2,0);
\draw[-{Latex[length=3mm]}] (1.56366, 1.24698) arc[start angle=38.5714, end angle=10.6913,radius=2cm];
\draw[-{Latex[length=3mm]}] (1.94986, -0.445042) arc[start angle=-12.8571, end angle=-47.1658,radius=2cm];
\draw[-{Latex[length=3mm]}] (-0.867767, -1.80194) arc[start angle=-115.714, end angle=-150.023,radius=2cm];
\draw[-{Latex[length=3mm]}] (-2, 0) arc[start angle=-180, end angle=-207.88,radius=2cm];
\node (0) at (0.8,0.8) {$\mathcal{L}_1$};
\node (1) at (0.8,-0.8) {$\mathcal{L}_2$};
\node (2) at (-0.8,-0.8) {$\mathcal{L}_3$};
\node (3) at (-0.8,0.8) {$\mathcal{L}_4$};
\node (4) at (2.73641, 2.18221) {$1$};
\node (5) at (3.41225, -0.778823) {$2$};
\node (6) at (1.51859, -3.15339) {$3$};
\node (7) at (-1.51859, -3.15339) {$4$};
\node (8) at (-3.41225, -0.778823) {$5$};
\node (9) at (-2.73641, 2.18221) {$6$};
\node (10) at (0,3.5) {$7$};
\node (11) at (2.43732, 0.556302) {$y_2$};
\node (12) at (1.95458, -1.55872) {$y_3$};
\node (13) at (0, -2.5) {$y_4$};
\node (14) at (-1.95458, -1.55872) {$y_5$};
\node (15) at (-2.43732, 0.556302) {$y_6$};
\node (16) at (-1.08471, 2.25242) {$y_7$};
\node (17) at (1.08471, 2.25242) {$y_1$};
\end{tikzpicture}
\caption{${\cal{G}}_7$}
\end{subfigure}
\caption{Six- and seven-point graphs that are equivalent to the ziggurat graphs shown in Fig.~\ref{fig:one}(c) and (d) under the graphical moves considered in~\cite{Prlina:2018ukf}. The labeling on (b) will be used in Sec.~\ref{sec:main}. The arrows indicate the four propagators carrying loop momentum $\ell_a$ (adjacent to the corresponding $\mathcal{L}_a$).}
\label{fig:two}
\end{figure}
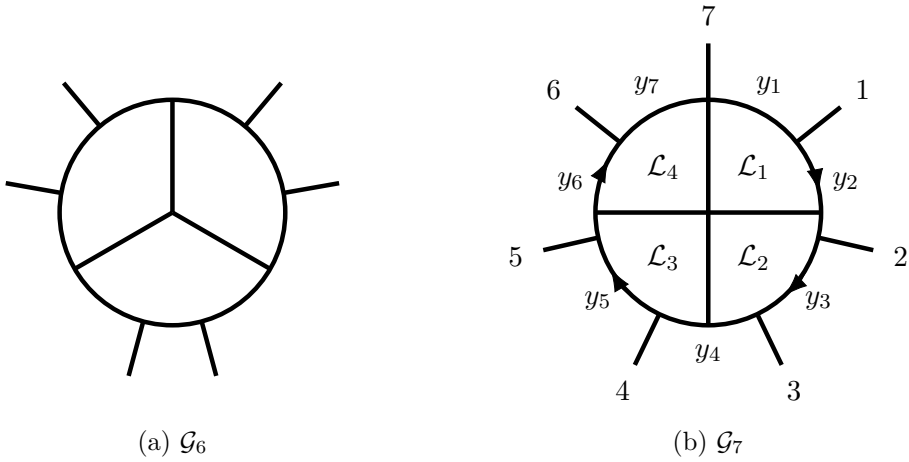

In this paper we compute the leading (first-type) Landau singularities of the four-loop seven-point graph ${\cal{G}}_7$ shown in Fig.~\ref{fig:two}(b). The motivation for looking at this particular graph stemmed from the analysis of~\cite{Prlina:2018ukf}, which argued that in massless planar theories, the locus of solutions to the Landau equations is invariant under certain simple graphical moves, in particular including the wye-delta (also called star-triangle) transformation, and that any planar graph can be reduced via these transformations to a ``ziggurat'' graph of the type shown in Fig.~\ref{fig:one}. However, it was noted later in~\cite{Lippstreu:2023oio} that for certain graphs there can exist branches of solutions to the Landau equations that are not preserved under the wye-delta transformation. Although this realization certainly reduces the impact of determining the Landau singularities of ${\cal{G}}_7$, there are still several reasons why it may be useful to do so.

For one thing, as mentioned above, the heptagon bootstrap begins with the assumption that the 49-letter heptagon symbol alphabet of~\cite{Drummond:2014ffa} describes the complete set of singularities for all seven-point amplitudes in planar $\mathcal{N}=4$ SYM theory. This assumption has enabled the computation of these amplitudes through to four-loops~\cite{Dixon:2020cnr,Drummond:2018caf,Drummond:2014ffa}. However, there remains the logical possibility that at higher loop order, these amplitudes may exhibit singularities not contained in this presumed set, so it remains interesting to scour the singularities of various massless, planar seven-point Feynman integrals to see if there might be \emph{any} beyond those associated to the presumed heptagon alphabet. This question is of particular importance for the purpose of bootstrapping or analyzing amplitudes or integrals in more general theories, where one might worry that the singularity structure might be vastly more complicated without the special cancellations present in SYM theory. We are not aware of any previous Landau analysis, numerical or analytical, of a massless four-loop seven-point integral, so basic questions regarding the mathematical structure of the potential singularities of such objects remains unknown. Viewed in that light, it is an interesting and non-trivial finding of our analysis that $\mathcal{G}_7$'s leading first-type singularities are actually quite tame; they are all within the heptagon alphabet. 

In Sec.~\ref{sec:two} we review the Landau equations, and in particular their formulation in momentum twistor space where on-shell conditions can be solved via simple geometric considerations in many nontrivial examples. As illustrative examples we discuss in detail how to determine the singularities of the four- and five-point ziggurat graphs before reviewing the six-point graph ${\cal{G}}_6$ from~\cite{Prlina:2018ukf}. In Sec.~\ref{sec:main} we find the leading singularities of ${\cal{G}}_7$, which are all consistent with the heptagon symbol alphabet. The graph has numerous subgraphs that need to be analyzed separately, which we defer to the companion paper~\cite{Lippstreu:2023oio}.

\newpage

\section{Warm-up}
\label{sec:two}

In~\cite{Landau:1959fi} Landau showed that a Feynman integral can have singularities only if certain polynomial equations are satisfied. In this section we review (largely following~\cite{Prlina:2017azl}) the formulation of the Landau equations in momentum twistor space. A significant advantage of working in momentum twistor space~\cite{Hodges:2009hk} is that some of the Landau equations (the on-shell conditions) can be solved analytically, in many non-trivial cases (such as the showcase of this paper: $\mathcal{G}_7$), via simple geometric considerations (see for example~\cite{Arkani-Hamed:2010pyv} for some simpler examples). We also review the role of the ziggurat graphs studied in~\cite{Prlina:2018ukf}.

\subsection{Landau equations in momentum twistor space}

In momentum twistor space each external leg of a planar $n$-point graph is associated to a point $Z_i$ in $\mathbb{P}^3$ (with an implied cyclic ordering of $1,\ldots,n$) and the external face bounded by legs $Z_i$, $Z_{i+1}$ is associated to the line containing those points, which we denote by $(Z_i, Z_{i+1})$ or simply $(i\,i{+}1)$. Momentum twistors are related to the standard Mandelstam variables by
\begin{align}
(p_i + p_{i+1} + \cdots + p_j)^2 = \frac{\braket{i{-}1\,i\,j\,j{+}1}}{\braket{i{-}1\,i\,\mathcal{I}}\braket{j{-}1\,j\,\mathcal{I}}}
\end{align}
where $\braket{ABCD}$ denotes the $4 \times 4$ determinant of the homogeneous coordinates of four points in $\mathbb{P}^3$ and $\mathcal{I} = (\mathcal{I}_1, \mathcal{I}_2)$ represents the ``line at infinity'', the choice of which necessarily breaks dual conformal invariance. To put it another way, any quantities involving $\mathcal{I}$ must cancel out in SYM theory since there is no invariant notion of ``infinity'' in momentum space. For applications to scattering amplitudes $\mathcal{I}$ is conventionally chosen so that $\braket{i\, j\, \mathcal{I}}=\braket{i\, j}$ coincides with the bracket of the two-component spinor helicity variables. We refer readers to~\cite{Mason:2009qx} for a more detailed introduction to (momentum) twistor geometry. 

Each internal face of a graph is associated to a line $\mathcal{L}_\ell$ representing the loop integration degrees of freedom. Often we parameterize $\mathcal{L}_\ell$ as $(A_\ell, B_\ell)$ for a choice of two distinct points on the line. We label a propagator bounded by two faces $(A,B)$ and $(C,D)$ by $\braket{ABCD}$. While the numerical value of this quantity is ambiguous, depending on the choices of representative points $(A,B)$ and $(C,D)$ on the two lines, it vanishes if and only if the two lines intersect, which is all that we will be interested in: this corresponds to the propagator going on-shell.

Consider an $L$-loop planar graph with $p$ propagators labeled $f_1, \ldots, f_p$. The Landau equations come in two types. First we have the on-shell conditions
\begin{align}
f_J = 0 \qquad J=1,2,\ldots,p\,,
\label{eq:onshell}
\end{align}
(a solution of which we naturally call a cut), and next we have the Kirchhoff conditions
\begin{align}
\sum_{J=1}^p \alpha_J \frac{\partial f_J}{\partial c_A} = 0 \qquad A=1,\ldots,4L\,,
\label{eq:kirch}
\end{align}
where the $\alpha_J$'s are Feynman parameters and the $c_A$ stand for $4L$ independent variables in terms of which we choose to parameterize the $L$ loop momenta. For the sake of definiteness we can choose for example
\begin{align}
\mathcal{L}_1 = \begin{pmatrix}
1 & 0 & c_1 & c_2 \\
0 & 1 & c_3 & c_4
\end{pmatrix},
\qquad
\mathcal{L}_2 = \begin{pmatrix}
1 & 0 & c_5 & c_6 \\
0 & 1 & c_7 & c_8
\end{pmatrix},
\qquad
\text{etc,}\label{eq:gauge}
\end{align}
where we use the $GL(2)$-invariance of the description of each line $\mathcal{L}_i$ to bring it to the above gauge-fixed form.  We always exclude trivial solutions to~(\ref{eq:kirch}) having all $\alpha_J = 0$. The singularities encoded in the Landau equations~(\ref{eq:onshell}) and~(\ref{eq:kirch}) are sometimes called \emph{first-type} singularities, in contrast to \emph{second-type} singularities which arise from pinch singularities at infinite loop momentum~\cite{ELOP,Fairlie:1962-1,Fairlie:1962-2}. The latter commonly arise from triangle subdiagrams; see for example the discussion in Sec.~V of~\cite{Dennen:2015bet}. Henceforth whenever we talk about Landau singularities, we only mean first-type singularities.

The explicit evaluation of Feynman integrals that are UV and/or IR divergent requires a regularization procedure. All of the integrals we consider are UV finite, and our methods are suitable for treating IR divergences via dimensional regularization. Singularities appearing in terms that are divergent, finite, or vanishing as the dimensional regulator is taken to zero are all detected by solving the Landau equations.

To find the Landau singularities of any given graph it is necessary to analyze the Landau equations for the full graph itself as well as for any subgraph that can be obtained by contracting any subset of its propagators. Because this is tantamount to setting various $\alpha$'s to zero, the on-shell conditions~(\ref{eq:onshell}) are traditionally written as $\alpha_J f_J = 0$ to emphasize that one can consider the two cases $\alpha_J = 0$ or $f_J = 0$ separately. However for massless diagrams, it is common (as we will see below) to have solutions with both $\alpha_J = 0$ and $f_J = 0$ for one or more $J$'s. For bookkeeping purposes we find it more convenient to demand that all propagators of a given graph must be put on-shell, and then remember to analyze all possible subgraphs separately. The recent papers~\cite{Fevola:2023fzn, Fevola:2023kaw} have highlighted that certain solutions to Landau equations can be missed if one does not allow for the possibility that the various Feynman parameters may approach zero at different rates. It would be very interesting to investigate this possibility in the context of our applications; until that is done, our results can only be interpreted as yielding a subset of potential kinematic singularities. 

Note that (other than excluding trivial solutions) we are never interested in the values of the $\alpha$'s (or $c$'s), only in the binary question: what constraints must the external $Z_i$ satisfy in order for nontrivial solutions of the Landau equations to exist? This is the locus of (potential) singularities for any Feynman integral involving the propagators indicated in the graph under consideration. (Specific choices of numerator factors in a Feynman integral may conspire to cancel some of the potential singularities; the Landau equations are manifestly blind to numerators and only care about the propagator structure, which is encoded in the graph topology.)

\subsection{Ziggurat graphs}
\label{sec:zig}

In~\cite{Dennen:2016mdk,Prlina:2018ukf} it was argued that when all propagators are \emph{massless}, the locus of solutions to the Landau equations associated to a graph is invariant under the graphical moves familiar from electrical circuit theory: series reduction, parallel reduction, and most importantly the wye-delta transform.  The problem of classifying all \emph{planar} graphs under these graphical moves has been solved in terms of what were called ``ziggurat'' graphs in~\cite{Prlina:2018ukf}. Specifically, any planar $n$-point graph is equivalent to the $n$-point ziggurat graph or a minor thereof. A minor of a graph is any graph that can be obtained by any combination of edge contractions or edge deletions. (Singularities associated to a given initial graph are called its \emph{leading} singularities, while those associated to a minor are called \emph{subleading} singularities.)

Combining these arguments would therefore suggest that the Landau singularities of any $n$-point massless planar Feynman integral are therefore a subset of those of the $n$-point ziggurat graph and its minors. However, it was shown in~\cite{Lippstreu:2023oio} that for graphs with a completely internal 3-vertex, there can exist branches of solutions to the Landau equations which are absent from the graph where the vertex is wye-delta transformed into a triangle.

\subsection{The four-point ziggurat}

In order to demonstrate the procedure of solving Landau equations in momentum twistor space we begin with the massless box graph shown in Fig.~\ref{fig:one}(a). The on-shell conditions are
\begin{align}
\braket{\mathcal{L}12} = \braket{\mathcal{L}23} = \braket{\mathcal{L}34} = \braket{\mathcal{L}41} = 0\,,
\end{align}
where we use the shorthand $\braket{\mathcal{L} i j} = \braket{\mathcal{L} Z_i Z_j}$. These admit two distinct solutions
\begin{align}
\mathcal{L} = (Z_1, Z_3) \qquad \text{or} \qquad \mathcal{L} = (Z_2, Z_4)\,.
\end{align}
The Kirchhoff conditions take the form of a $4 \times 4$ matrix multiplying $(\alpha_1\,\alpha_2\,\alpha_3\,\alpha_4)$ to give zero. Nontrivial solutions exist only when the determinant of this matrix, which evaluates to $\braket{1234}^2$ on either of the two on-shell solutions, vanishes. This agrees with the usual momentum space analysis which reveals that (in terms of $s = (p_1 + p_2)^2$, $t = (p_2 + p_3)^2$) the massless box integral has leading Landau singularities only when
\begin{align}
s\, t = \frac{\braket{1234}^2}{\braket{12\mathcal{I}} \braket{23\mathcal{I}} \braket{34\mathcal{I}} \braket{41\mathcal{I}}} = 0\,.
\end{align}
Interestingly we don't see any sign of IR divergences in the leading Landau singularity.  However if we contract (for example) the fourth edge, the three remaining on-shell conditions then admit two one-parameter families of solutions
\begin{align}
\mathcal{L} = (Z_2, \alpha Z_3 + (1 - \alpha) Z_4) \qquad \text{or} \qquad \mathcal{L} = (Z_3, \alpha Z_1 + (1 - \alpha) Z_2)\,.
\label{eq:triangle}
\end{align}
The Kirchhoff conditions are no longer equivalent to a vanishing determinant since they are non-linear in the remaining variables ($\alpha_1, \alpha_2, \alpha_3, \alpha$). Nevertheless it is easy to check that the Landau equations admit the nontrivial solution $\mathcal{L} = (Z_2,Z_3)$, $\alpha_1 = \alpha_3 = 0$ for arbitrary external kinematics $Z_i$. We interpret solutions that exist for all kinematics as signaling the presence of IR singularities, arising from the soft/collinear region of loop momentum space. Going forward we are interested in classifying solutions of the Landau equations that exist only on codimension-one surfaces in the space of external kinematics, since these determine the locus of (potential) branch points (``branch surfaces'', really) of an integral. (See~\cite{Bourjaily:2022vti} for an interesting recent discussion of higher codimension singularities.)

Altogether, after checking the Landau equations for all (triangle, bubble, or tadpole) subdiagrams one can obtain from the box by any combination of edge contractions, and discarding all solutions corresponding to IR singularities, one finds no additional singularities beyond the one at $\braket{1234}=0$ present already in the box's leading singularity.

\subsection{The five-point ziggurat}
\label{sec:five}

We label the external edges of the five-point ziggurat graph shown in Fig~\ref{fig:one}(b) with $Z_1$ on the lower right corner, increasing in clockwise order, and we label the lower (upper) loop with $\mathcal{L}_1$ ($\mathcal{L}_2$) respectively. It is well-known that the seven on-shell conditions (i.e., the double box heptacut) admit six distinct one-parameter families of solutions (see for example~\cite{Caron-Huot:2012awx} for a nice discussion):
\begin{alignat*}{2}
\mathcal{L}_1 &= (Z_1, \alpha Z_2 + (1{-}\alpha) Z_3) && \mathcal{L}_1 = (Z_1, \alpha Z_2 + (1{-}\alpha) Z_3) \\
\mathcal{L}_2 &= (Z_1, Z_4) \qquad && \mathcal{L}_2 = (Z_5, (\alpha \braket{1245} + (1{-}\alpha) \braket{1345}) Z_3 - \alpha \braket{1235} Z_4) \\[10pt]
\mathcal{L}_1 &= (Z_1, \braket{1345} Z_2 - \braket{1245} Z_3) \qquad && \mathcal{L}_1 = (Z_2, \alpha Z_1 + (1{-}\alpha) Z_5) \\
\mathcal{L}_2 &= (Z_4, \alpha Z_1 + (1{-}\alpha) Z_5) && \mathcal{L}_2 = (Z_5, \braket{1245} Z_3 - \braket{1235} Z_4) \\[10pt]
\mathcal{L}_1 &= (Z_2, Z_5) && \mathcal{L}_1 = (Z_2, \alpha Z_1 + (1{-}\alpha) Z_5) \\
\mathcal{L}_2 &= (Z_5, \alpha Z_3 + (1{-}\alpha) Z_4) && \mathcal{L}_2 = (Z_4, \alpha Z_1 + (1{-}\alpha) Z_5)
\end{alignat*}

Plugging (for example) the first on-shell solution into the Kirchhoff conditions gives eight equations in eight variables (the seven Feynman parameters and the on-shell parameter $\alpha$). There are seven non-trivial solutions which are at most codimension one in the external kinematics. One of these solutions has $\alpha = 1$ (so that $\mathcal{L}_1 = (Z_1,Z_2)$) and all Feynman parameters vanishing except for the one associated to the propagator $\braket{\mathcal{L}_1 1 2}$. This solution exists for all external kinematics and we interpret as an IR singularity, as discussed in the previous section.  In addition to such uninteresting solutions, there are other solutions that only exist when $\braket{1345}\braket{1245}\braket{1235} = 0$.

After repeating this analysis for the other on-shell solutions, and considering also all cyclic relabelings of the ziggurat graph, one finds that leading Landau singularities can exist when $\braket{i\,i{+}1\,i{+}2\,i{+}3} = 0$ for some $i$. In momentum space this corresponds to $(p_i + p_{i+1})^2 = s_{i,i+1} = 0$. As in the four-point case, a thorough analysis of the Landau equations for every minor of the ziggurat graph Fig.~\ref{fig:one}(b) reveals no additional singularities beyond those of the form $\braket{i\,i{+}1\,i{+}2\,i{+}3}$ already present in the ziggurat's leading singularities.

Let us pause here to emphasize that there is no tension between this result and the fact that massless planar five-point Feynman integrals with more complicated singularities are certainly known. For example, the two-loop master integrals relevant to five-point functions in massless planar QCD have a 26-letter symbol alphabet~\cite{Gehrmann:2015bfy} that indicates the presence of branch point singularities at $s_{i,i+1} + s_{i+1,i+2} = 0$ (and other even more complicated functions of the Mandelstam variables). The fact that $s_{i,i+1} + s_{i+1,i+2} = 0$ cannot be expressed without introducing a choice of infinity twistor $\mathcal{I}$ is a giveaway that these more complicated letters arise as second-type singularities, and would never be detected by our momentum twistor analysis.

\subsection{\texorpdfstring{The six-point graph ${\cal{G}}_6$}{The six-point graph G6}}

The six-point graph ${\cal{G}}_6$ was analyzed in~\cite{Prlina:2018ukf}. This graph has the advantage of having precisely four times as many propagators as loops. This means that like in the four-point case, the solutions to the on-shell conditions are discrete (here there are 16 instead of 2) and the Kirchhoff conditions can be expressed as a determinant (here it is $12 \times 12$ instead of $4 \times 4$). Also like in the four-point case, the analysis of the leading singularity is not clouded by the need to isolate and discard solutions corresponding to IR singularities. When evaluated on any one of the on-shell solutions, the Kirchhoff determinant factors into a product of various four-brackets $\braket{i\,j\,k\,l}$. By scanning over all 16 solutions, and considering all independent cyclic images of the graph, one encounters all $\binom{6}{4} = 15$ distinct four-brackets.

This analysis in~\cite{Prlina:2018ukf} was interpreted as bolstering the expectation---consistent with all results available to date, which now extends to seven loop order~\cite{Caron-Huot:2019vjl}---that the 15 four-brackets constitute the symbol alphabet for all six-point amplitudes in SYM theory, to any loop order. However, thanks to the results of~\cite{Lippstreu:2023oio} we know that this argument is incorrect, even if its conclusion turns out to be correct.

\section{\texorpdfstring{The seven-point graph ${\cal{G}}_7$}{The seven-point graph G7}}
\label{sec:main}

\begin{figure}
\centering
\includegraphics[width=6.125truein]{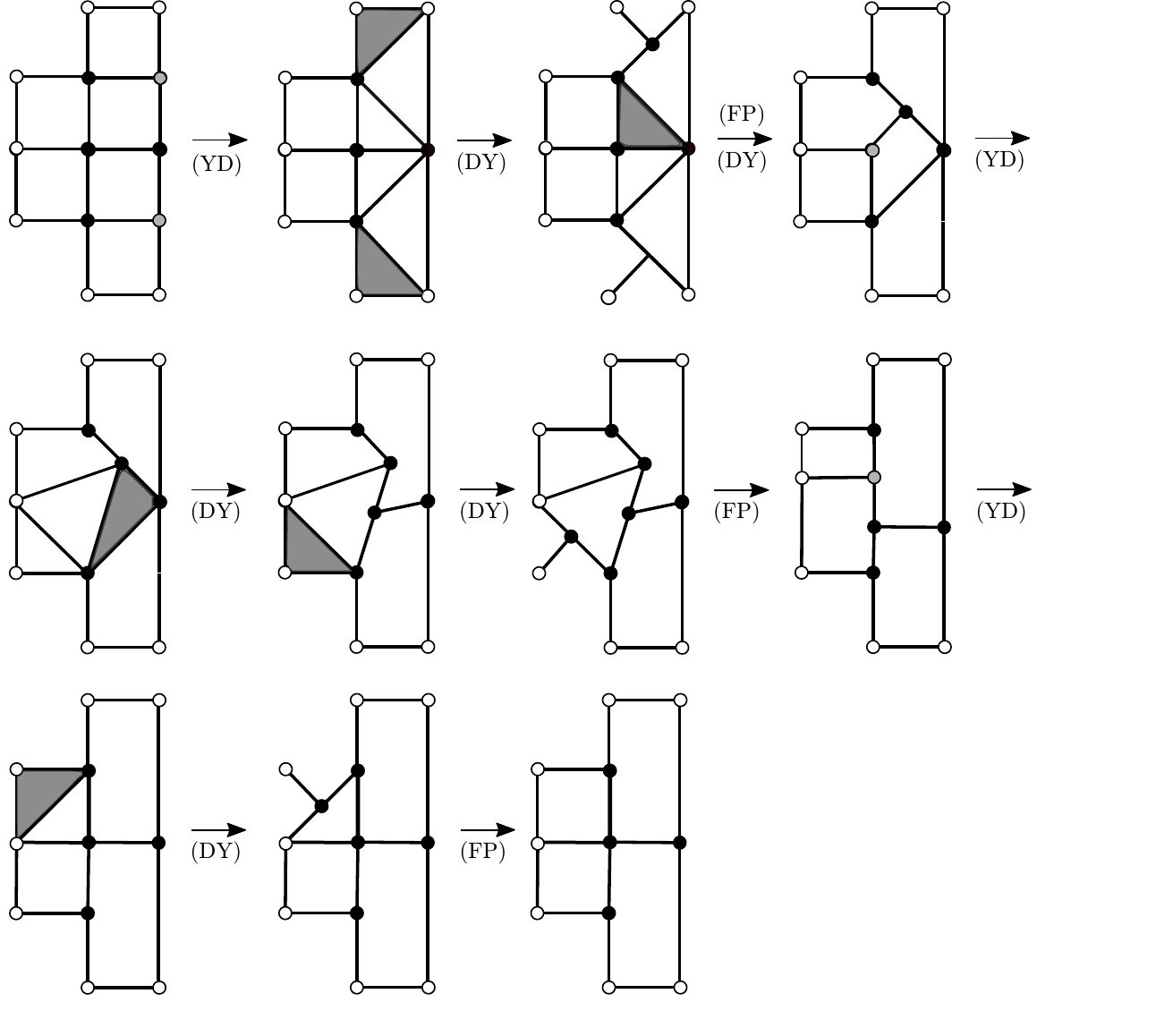}
\caption{A sequence of graphical moves (see~\cite{Prlina:2018ukf}) that transforms the seven-point ziggurat graph Fig.~\ref{fig:one}(d) into the wheel graph Fig.~\ref{fig:two}(b). YD indicates wye-delta transformation(s) on the node(s) shaded in grey; DY indicates delta-wye transformations(s) on the triangle(s) shaded in grey, and FP indicates a trivial contraction of external edges.}
\label{fig:three}
\end{figure}

In this section we outline the calculation of the (leading) Landau singularities of the seven-point graph ${\cal{G}}_7$ shown in Fig.~\ref{fig:two}(b). We will see that this analysis is computationally similar to the five-point calculation reviewed in Sec.~\ref{sec:five}. Our original motivation for studying this graph is that it can be obtained from the seven-point ziggurat graph in Fig.~\ref{fig:one}(d) by the sequence of graphical moves shown in Fig.~\ref{fig:three}. However, now that we know that the wye-delta transform is not faithful to the locus of Landau singularities~\cite{Lippstreu:2023oio}, our analysis of ${\cal{G}}_7$ stands on its own as a showcase of Landau-solving technology, and an attempt to begin gathering data about the kinds of singularities that can appear in such integrals.

\subsection{Solving the on-shell conditions}

We now turn to the first set of Landau equations--the on-shell conditions. Since the graph Fig.~\ref{fig:two}(b) has 14 propagators, and there are 16 degrees of freedom in the four loop momenta, we expect solutions to come in 2-parameter families.

To organize the calculation we first consider the two propagators
\begin{align}
\braket{\mathcal{L}_171} = \braket{\mathcal{L}_112} = 0
\end{align}
which are of ``one-mass bubble'' type, referring to the standard terminology (see for example Tab.~1 of~\cite{Prlina:2017azl}), and have two solutions: one for which the line $\mathcal{L}_1$ passes through the point $Z_1$, and one for which $\mathcal{L}_1$ lies in the plane $\bar{1}$. (Here we use the standard notation $\bar{i} = (i{-}1\,i\,i{+}1)$, with $(ijk)$ denoting the plane containing $Z_i$, $Z_j$ and $Z_k$.) Since the Landau equations are parity invariant, the parity conjugate of any solution is again a solution, so it suffices to consider only the case where $\mathcal{L}_1$ passes through $Z_1$ and obtain the remaining solutions by parity conjugation. For example, note that parity acts on a 4-bracket contraction as
\begin{gather}
    \text{Parity:  }\braket{ABCD}\mapsto\braket{(\bar{A}\cap \bar{B}),(\bar{C}\cap \bar{D})}.
\end{gather}
 Altogether we find a total of 40 two-dimensional solutions to the on-shell equations. Half of these solutions are listed in Tables~\ref{tab:firstonshellsol} and~\ref{tab:secondonshellsol}; the rest are their parity conjugates.

\begin{table}[tbp]
\centering
\small
\begin{tabular}{|c|ll|c|}
\hline
\begin{tabular}[c]{@{}c@{}}${}$\\ \# \\ ${}$\end{tabular} & \multicolumn{2}{c|}{\begin{tabular}[c]{@{}c@{}}${}\hspace{-2cm}\begin{pmatrix} \mathcal{L}_1 && \mathcal{L}_4\\ \mathcal{L}_2 && \mathcal{L}_3\end{pmatrix}\qquad{}$\end{tabular}} & \begin{tabular}[c]{@{}c@{}}Letter\\ Classes \end{tabular}\\ \hline

\multirow{2}{*}{1} & $(Z_1,l_{24}(\alpha))$ & $(Z_6, \beta Z_1+(1-\beta)l_{24}(\alpha))$  & \multirow{2}{*}{$b_{0},b_{2},b_{4}$ } \\
 & $(Z_2,Z_4)$ & $(Z_4,Z_6)$ &\\ \hline
 %%%%%%%%%%%%%%%%%%%%%%%%
 
\multirow{2}{*}{2} & $(Z_1,l_{24}(\alpha))$ & $(Z_6,(Z_1,l_{24}(\alpha))\cap \bar{5})$ & $b_{0},b_{1},b_{2},$ \\
 & $(Z_2,Z_4)$ & $(Z_4,l_{56}(\beta))$ & $b_{4}$\\ \hline

  %%%%%%%%%%%%%%%%%%%%%%%%
\multirow{2}{*}{3} & $(Z_1,P)$ & $(Z_6, (1,P)\cap(356))$ & \multirow{2}{*}{$b_{0},b_{1},b_{3}$} \\
 & $(Z_3,Z_1)$ & $(Z_5,Z_3)$ &\\ \hline

  %%%%%%%%%%%%%%%%%%%%%%%%
  
\multirow{2}{*}{4} & $(Z_1,l_{23}(\alpha))$ & $(Z_6,(l_{23}(\alpha),Z_1)\cap (356))$ & \multirow{2}{*}{$b_{0},b_{1},b_{3}$} \\
 & $(Z_3,l_{12}(\beta))$ & $(Z_5,Z_3)$ & \\ \hline
 %%%%%%%%%%%%%%%%%%%%%%%%%%%%%%

\multirow{2}{*}{5} & $(Z_1,Z_2(1{-}\beta)+\beta l_{34}(\alpha))$ & $\Big(Z_6,\big(Z_1,Z_2(1{-}\beta)+\beta l_{34}(\alpha)\big)\cap\big(l_{34}(\alpha),Z_5,Z_6\big)\Big)$ & $b_{0},b_{1},b_{2},$ \\
 & $(Z_2,l_{34}(\alpha))$ & $(Z_5,l_{34}(\alpha))$ & $b_{3},b_{5}$\\ \hline

\multirow{2}{*}{6} & $(Z_1,l_{23}(\alpha))$ & $(Z_6,\beta Z_1+(1-\beta)l_{23}(\alpha))$ & \multirow{2}{*}{$b_{0},b_{3}$}\\
 & $(Z_3,(12)\cap\bar{4})$ & $(Z_5,(34)\cap (Z_5,Z_6,\beta Z_1+(1-\beta)l_{23}(\alpha)))$ &  \\ \hline

\multirow{2}{*}{7} & $(Z_1,\beta Z_2+(1-\beta)l_{34}(\alpha))$ & $(Z_6,(Z_1,\beta Z_2+(1-\beta)l_{34}(\alpha))\cap \bar{5})$ & \multirow{2}{*}{$b_{0},b_{1},b_{2}$} \\
 & $(Z_2,l_{34}(\alpha))$ & $(Z_4,(56)\cap\bar{3})$ & \\ \hline

\multirow{2}{*}{8} & $(Z_1,P)$ & $(Z_6,(Z_1,P)\cap\bar{5})$ & $b_{0},b_{1},b_{3},$ \\
 & $(Z_3,Z_1)$ & $(Z_4,(56)\cap(134))$ & $b_{4}$ \\ \hline

\multirow{2}{*}{9} & $(Z_1,l_{23}(\alpha))$ & $(Z_6,\beta Z_1+(1-\beta)l_{23}(\alpha))$ & $b_{0},b_{2},b_{3},$ \\
 & $(Z_3,(12)\cap (346))$ & $(Z_4,Z_6)$ & $b_{4}$\\ \hline

\multirow{2}{*}{10} & $(Z_1,l_{23}(\alpha))$ & $(Z_6,(Z_1,l_{23}(\alpha))\cap \bar{5})$  & $b_{0},b_{1},b_{2},$\\
 & $(Z_3,l_{12}(\beta))$ & $(Z_4,(56)\cap(l_{12}(\beta),Z_3,Z_4))$ & $b_{3},b_{4}$ \\ \hline
\end{tabular}
\caption{The first ten solutions to the on-shell conditions associated to Fig.~\ref{fig:two}(b). These solutions all have the line $\mathcal{L}_1$ passing through $Z_1$ and the line $\mathcal{L}_4$ passing through $Z_6$. Here $\bar{i}$ indicates the plane $(i{-}1\,i\,i{+}1)$, $l_{ij}(\alpha)=\alpha Z_i+(1-\alpha)Z_j$ is a point on the line $(ij)$, and $P$ denotes an arbitrary twistor. Each solution has two degrees of freedom, manifested in most cases by the arbitrary parameters $\alpha$ and $\beta$. The arbitrary point $P$ in solutions \#3 and \#8 has three degrees of freedom, but shifting $P$ in the direction of $Z_1$ leaves the solution unchanged so there are effectively only two degrees of freedom. The third column indicates which symbol letters (see main text) appear as Landau singularities for each cut.}
\label{tab:firstonshellsol}
\end{table}

\begin{table}[tbp]
\centering
\begin{tabular}{|c|ll|c|}
\hline
\begin{tabular}[c]{@{}c@{}}${}$\\ \#\\ ${}$\end{tabular} & \multicolumn{2}{c|}{\begin{tabular}[c]{@{}c@{}}${}\begin{pmatrix} \mathcal{L}_1 && \mathcal{L}_4\\ \mathcal{L}_2 && \mathcal{L}_3\end{pmatrix}\qquad{}$\end{tabular}} & \begin{tabular}[c]{@{}c@{}}Letter\\ Classes \end{tabular} \\ \hline
\multirow{2}{*}{11} & $(Z_1,l_{24}(\alpha))$ & $\bar{6}\cap(Z_1,l_{24}(\alpha),l_{56}(\beta))$ & \multirow{2}{*}{$b_0,b_1,b_2$}\\
 & $(Z_2,Z_4)$ & $(Z_4,l_{56}(\beta))$ &  \\ \hline
  %%%%%%%%%%%%%%%%%%%%%%%%%%%%%%%%%%%%%%
 %%%%%%%%%%%%%%%%%%%%%%%%%%%%%%%%%%%%%
\multirow{2}{*}{12} & $(Z_1,(23)\cap (Z_1,Z_5,P))$ & $\bar{6}\cap (Z_1,Z_5,P)$  & \multirow{2}{*}{$b_0,b_1,b_3$}\ \\
 & $(Z_3,l_{12}(\alpha))$ & $(Z_5,Z_3)$ & \\ \hline
 %%%%%%%%%%%%%%%%%%%%%%%%%%%%%%%%%%%%%%
 %%%%%%%%%%%%%%%%%%%%%%%%%%%%%%%%%%%%%
\multirow{2}{*}{13} & $(Z_1,P)$ & $\bar{6}\cap (Z_1,Z_5,P)$ & \multirow{2}{*}{$b_0,b_1,b_3$} \\
 & $(Z_3,Z_1)$ & $(Z_5,Z_3)$ & \\ \hline
  %%%%%%%%%%%%%%%%%%%%%%%%%%%%%%%%%%%%%%
 %%%%%%%%%%%%%%%%%%%%%%%%%%%%%%%%%%%%%
\multirow{2}{*}{14} & $(Z_1,l_{2,(34)\cap \bar{6}}(\alpha))$ & $\bar{6}\cap (Z_1,l_{2,(34)\cap\bar{6}}(\alpha),P)$ &\multirow{2}{*}{$b_0,b_3,b_5$} \\
 & $(Z_2,(34)\cap\bar{6})$ & $(Z_5,(34)\cap\bar{6})$ &  \\ \hline
  %%%%%%%%%%%%%%%%%%%%%%%%%%%%%%%%%%%%%%
 %%%%%%%%%%%%%%%%%%%%%%%%%%%%%%%%%%%%%
\multirow{2}{*}{15} & $(Z_1,(2,l_{34}(\alpha))\cap(Z_1,Z_5,P))$ & $\bar{6}\cap(Z_1,Z_5,P)$ & \multirow{2}{*}{$b_0,b_1,b_3$} \\
 & $(Z_2,l_{34}(\alpha))$ & $(Z_5,l_{34}(\alpha))$ & \\ \hline
  %%%%%%%%%%%%%%%%%%%%%%%%%%%%%%%%%%%%%%
 %%%%%%%%%%%%%%%%%%%%%%%%%%%%%%%%%%%%%
\multirow{2}{*}{16} & $(Z_1,l_{23}(\alpha))$ & $\bar{6}\cap (Z_1,l_{23(\alpha)},P)$ & \multirow{2}{*}{$b_0,b_3$}\\
 & $(Z_3,(12)\cap\bar{4})$ & $(Z_5,(34)\cap \bar{6})$ &  \\ \hline
  %%%%%%%%%%%%%%%%%%%%%%%%%%%%%%%%%%%%%%
 %%%%%%%%%%%%%%%%%%%%%%%%%%%%%%%%%%%%%
\multirow{2}{*}{17} & $(Z_1,l_{23}(\alpha))$ & $\bar{6}\cap(Z_1,l_{23}(\alpha),5)$ & \multirow{2}{*}{$b_0,b_1,b_3$} \\
 & $(Z_3,(12)\cap\bar{4})$ & $(Z_5,l_{34}(\alpha))$ & \\ \hline
  %%%%%%%%%%%%%%%%%%%%%%%%%%%%%%%%%%%%%%
 %%%%%%%%%%%%%%%%%%%%%%%%%%%%%%%%%%%%%
\multirow{2}{*}{18} & $(Z_1,\beta Z_2+(1-\beta)l_{34}(\alpha))$ & $\bar{6}\cap\Big(Z_1,\beta Z_2+(1-\beta)l_{34}(\alpha),(56)\cap\bar{3}\Big)$ & \multirow{2}{*}{$b_0,b_1,b_2$}\\
 & $(Z_2,l_{34}(\alpha))$ & $(Z_4,(56)\cap\bar{3})$ & \\ \hline
  %%%%%%%%%%%%%%%%%%%%%%%%%%%%%%%%%%%%%%
 %%%%%%%%%%%%%%%%%%%%%%%%%%%%%%%%%%%%%
\multirow{2}{*}{19} & $(Z_1,l_{23}(\alpha))$ & $\bar{6}\cap\Big(Z_1,l_{23}(\alpha),(56)\cap(l_{12}(\beta),Z_3,Z_4)\Big)$ & $b_{0},b_{1},b_{2},$ \\
 & $(Z_3,l_{12}(\beta))$ & $\Big(Z_4,(56)\cap(l_{12}(\beta),Z_3,Z_4)\Big)$ & $b_{3},b_{4}$ \\ \hline
  %%%%%%%%%%%%%%%%%%%%%%%%%%%%%%%%%%%%%%
 %%%%%%%%%%%%%%%%%%%%%%%%%%%%%%%%%%%%%
\multirow{2}{*}{20} & $(Z_1,P)$ & $\bar{6}\cap \Big(Z_1,P,(56)\cap(134)\Big)$& $b_{0},b_{1},b_{3},$ \\
 & $(Z_3,Z_1)$ & $\Big(Z_4,(56)\cap(134)\Big)$  &$b_4$\\ \hline
\end{tabular}
\caption{The second ten solutions to the on-shell conditions associated to Fig.~\ref{fig:two}(b). These solutions all have the line $\mathcal{L}_1$ passing through $Z_1$ and the line $\mathcal{L}_4$ lying in the plane $\bar{6}$. Each solution has two degrees of freedom.}
\label{tab:secondonshellsol}
\end{table}

\subsection{Solving the Kirchhoff conditions}

When evaluated on any one of the on-shell solutions, the Kirchhoff conditions provide a system of 16 equations in 16 variables: the 14 Feynman parameters (which appear linearly) and two parameters associated to the cut (which in general appear nonlinearly). These equations are much more difficult to solve analytically than the on-shell conditions, though we report some very helpful intermediate results below. In general we find it necessary to adopt a ``numerical experimentation'' approach. Specifically, we populate the $4 \times 7$ matrix $Z$ of momentum twistors describing the external kinematics with 28 random integers, except for a single parameter ``$z$'' in some position. We then evaluate the Kirchhoff conditions on this one-parameter family of kinematic configurations, and find all solutions that exist only for certain values of $z$. Like the five-point calculation, this analysis is complicated by the fact that there are branches of solutions that exist for all values of $z$, which must be excluded. By iterating over all possible positions of the parameter $z$, and by repeating the calculation for many choices of random integer values for the other entries of $Z$, we can be sure that we have identified all codimension-one loci in kinematic space where the Landau equations admit solutions.

The last step is to make the connection between the Landau singularities found in this way and the vanishing of symbol letters; and specifically to test the expectation that the heptagon symbol alphabet captures the singularities of all seven-point amplitudes in SYM theory. The 49 symbol letters of the heptagon alphabet (see~\cite{Caron-Huot:2011zgw,Drummond:2014ffa,Dixon:2016nkn}) fall into seven classes under the $Z_i \to Z_{i+1}$ cyclic group. Let us denote the letters by
\begin{gather}
    b_{01}=\braket{1234},\qquad b_{11}=\braket{1256},\qquad b_{21}=\braket{1456},\qquad b_{31}=\braket{1236},\\
    b_{41}=\braket{1346},\qquad b_{51}=\braket{1(23)(45)(67)},\qquad b_{61}=\braket{1(34)(56)(72)},
\end{gather}
with $b_{ij}$ obtained from $b_{i1}$ by cyclically relabeling $Z_m \to Z_{m+j-1}$. Here $\braket{a(bc)(de)(fg)} = \braket{bade} \braket{cafg} - (b \leftrightarrow c)$. The letters of type $b_{0j}$, $b_{1j}$ and $b_{6j}$ are individually invariant under parity while the others are related under parity by
\begin{equation}
 b_{2,j} \leftrightarrow b_{3,j-1}\,, \qquad  b_{4,j}\leftrightarrow  b_{5,j-1}\,.
\end{equation}
In the third column of Tables~\ref{tab:firstonshellsol} and~\ref{tab:secondonshellsol} we indicate the families of symbol letters encountered for each of the on-shell solutions, using $b_i$ as shorthand for the cyclic family $\{b_{i1}, \ldots, b_{i7}\}$. Specifically, the Kirchhoff equations associated to a given cut admit (codimension-one) solutions only if the parameter $z$ takes a value that sets one or more symbol letters in an indicated family to zero. The union of singularities found in all cyclic relabelings of the original ziggurat graph comprises complete cyclic families. From the tables (and the parity conjugate cuts, which lead to analogous results with $b_2 \leftrightarrow b_3$ and $b_4 \leftrightarrow b_5$) we see that all heptagon symbol letters except for family $b_6$ are found as (leading) singularities of the seven-point graph Fig.~\ref{fig:two}(b). Singularities of type $b_6$ certainly appear in relaxations, as discussed in Sec.~\ref{sec:relaxations}.

Let us emphasize that the numerical approach we have outlined above relies on having a hypothesis for the set of singularities we are looking for. In our application, that hypothesis is the heptagon symbol alphabet, but the approach can be used more generally, for higher loop diagrams or for a larger number of external points, to detect whether any singularities occur outside of a proposed set of singularities. 

For each letter class displayed in a given cut solution in Tables~\ref{tab:firstonshellsol} and~\ref{tab:secondonshellsol} we have analysed the numerical solution which gave rise to that letter class and have been able to reconstruct an analytical solution to the Landau equations which produces that letter on the associated cut. This last analytical reconstruction step is done by inspecting the numerical solution and identifying which lines are soft and which lines are collinear. This informs an ansatz for what analytical solution the numerical solution corresponds to. We then check that the ansatz solution does indeed produce the associated letter, and also that it is not a solution which occurs for arbitrary external kinematics (which would then be associated to an IR-divergence instead). This last reconstruction step requires some familiarity with what analytical solutions to the Kirchhoff equations in momentum twistor space look like. We provide the reader with a prototypical example in the next section. 

\subsection{Some analytic details}

Let us provide some insight into solving the Kirchhoff equations analytically using momentum twistors, and in doing so demonstrate how to translate the twistor space results into kinematic configurations occurring in momentum space. The Kirchhoff equations associated to Fig.~\ref{fig:two}(b), written in momentum space read,
\begin{equation}
 \begin{aligned}
    \alpha_1(l_1-p_1)+\alpha_2l_1+\alpha_{11}(l_1+p_2-l_2)+\alpha_{14}(l_1-p_1-l_4-p_6-p_7)=0\,,\\
    \alpha_9l_4+\alpha_{10}(l_4+p_6)-\alpha_{13}(l_3+p_5-l_4)-\alpha_{14}(l_1-p_1-l_4-p_6-p_7)=0\,,\\
    \alpha_4l_2+\alpha_5(l_2+p_3)+\alpha_{12}(l_2+p_3-l_3+p_4)-\alpha_{11}(l_1+p_2-l_2)+\alpha_3(l_2-p_2)=0\,,\\
    \alpha_7l_3+\alpha_8(l_3+p_5)+\alpha_{13}(l_3+p_5-l_4)-\alpha_{12}(l_2+p_3-l_3+p_4)+\alpha_6(l_3-p_4)=0\,.\label{Kirchhoffp}
\end{aligned}   
\end{equation}
Each momentum $p_{a\dot{a}}$ in the graph is associated to four twistors $Z_A,Z_B,Z_C,Z_D$ (for external legs only three of the four twistors are distinct) with the explicit mapping given by
\begin{equation}
p_{a\dot{a}}(A,B,C,D)=\mathcal{I}^{\alpha\beta}\mathcal{I}_{\gamma\delta}\frac{\epsilon_{\beta}(\cdot,A,B,C)D^{\delta}-\epsilon_{\beta}(\cdot,A,B,D)C^{\delta}}{\braket{\mathcal{I}AB}\braket{\mathcal{I}CD}}\,,\label{eq:map p to twistor}
\end{equation}
where $\mathcal{I}^{\alpha\beta}\mathcal{I}_{\gamma\delta}$ denotes the infinity twistor and its dual, and the four twistors associated to a loop momentum occurring between the loop region $\mathcal{L}_i$ and the zone $y_j$ are $(\mathcal{L}_iZ_jZ_{j-1})$. For example $(l_1)_{a\dot{a}}=p_{a\dot{a}}(\mathcal{L}_1,Z_2,Z_1)$. The reader can verify that using~(\ref{eq:map p to twistor}) in~(\ref{Kirchhoffp}) and then going to the gauge (\ref{eq:gauge}) precisely reproduces the twistor space formulation of the Kirchhoff equations described in~(\ref{eq:kirch}).

Note that the first Kirchhoff equation in~(\ref{Kirchhoffp}) requires four four-dimensional vectors to be linearly dependent, which only occurs if their determinant vanishes. We can translate this vanishing determinant condition to $(a,\dot{a})$ indices via the relation
\begin{equation}
4i\e_{\mu\nu\rho\sigma}\sigma^{\mu}_{\dot{a}_1a_1}\sigma^{\nu}_{\dot{a}_2a_2}\sigma^{\rho}_{\dot{a}_3a_3}\sigma^{\sigma}_{\dot{a}_4a_4}= \e_{a_1a_2}\e_{\dot{a}_2\dot{a}_3}\e_{a_3a_4}\e_{\dot{a}_4\dot{a}_1}-\e_{\dot{a}_1\dot{a}_2}\e_{a_2a_3}\e_{\dot{a}_3\dot{a}_4}\e_{a_4a_1}\,,
\end{equation}
where $\sigma^{\mu}_{\dot{a}a}$ are the Pauli matrices. Using this we deduce that the first Kirchhoff equation admits solutions if either all of its Feynman parameters are zero, or if
\begin{gather}
      \text{Det}\big[l_1-p_1, l_1  , l_1+p_2-l_2 , l_1-p_1-l_4-p_6-p_7\big]\Big\vert_{\mathcal{L}_1=(Z_1,B)}=0\\
      \implies \braket{72\mathcal{L}_1}\braket{\mathcal{I}(\mathcal{L}_21)\cap (\mathcal{L}_41)}=0\label{firsdet}
\end{gather}
where the implication is understood only to hold on the support of our twenty cut solutions in Tables~\ref{tab:firstonshellsol} and~\ref{tab:secondonshellsol} which all take the form $\mathcal{L}_1=(Z_1,B)$. Similar constraints apply for the second Kirchhoff equation in~(\ref{Kirchhoffp}). For example on the support of the first ten cut solutions (Table~\ref{tab:firstonshellsol}) which have $\mathcal{L}_4=(Z_6,H)$ we deduce that either
\begin{gather}
\text{Det}\big[l_4,l_4+p_6,l_3+p_5-l_4,l_1-p_1-l_4-p_6-p_7\big]\Big\vert_{\mathcal{L}_4=(6,H)}=0\\
\implies\braket{57\mathcal{L}_4}\braket{\mathcal{I}(\mathcal{L}_36)\cap(\mathcal{L}_16)}=0\label{seconddet}
\end{gather}
or that all the Feynman parameters in the second Kirchhoff equation vanish.

Let us exemplify the kinematic configurations which solve the Landau equations by using the aforementioned results to find a solution to the Kirchhoff equations for the second cut in Table~\ref{tab:firstonshellsol}, which has $\mathcal{L}_1=(Z_1,l_{24}(\alpha))$. If we seek a solution where not all $\alpha$'s in the first Kirchhoff equation are zero, then we must necessarily satisfy~(\ref{firsdet}). A simple branch of solutions can be obtained by choosing $l_{24}=Z_2$ so that the first bracket in~(\ref{firsdet}) vanishes. Plugging this solution into the first Kirchhoff equation we find that this sets $(l_1)_{a\dot{a}}$ soft, in particular 
\begin{equation}
    (l_1)_{a\dot{a}}=p_{a\dot{a}}(Z_1,Z_2,Z_1,Z_2)=0\,,
\end{equation}
where we used that~(\ref{eq:map p to twistor}) vanishes when only two distinct twistors occur in its arguments.  Let us continue on this branch of solutions and seek a solution where not all Feynman parameters in the second Kirchhoff equation are zero, in which case~(\ref{seconddet}) must also be satisfied. A particular branch of solutions can be obtained by setting the first bracket in~(\ref{seconddet}) to zero,
\begin{equation}
\braket{576(1,2)\cap\bar{5}}=\braket{4567}\braket{1256}\overset{!}{=}0\,.\label{eq:vanish}
\end{equation}
By choosing either of the four brackets in~(\ref{eq:vanish}) to vanish we obtain two branches of solutions. Let us examine the $\braket{1256}=0$ branch. Using $\braket{1256}=0$ in the second Kirchhoff equation we find that $(l_{4})_{a\dot{a}}$ is set to zero,
\begin{align}
    (l_{4})_{a\dot{a}}&=p_{a\dot{a}}(\mathcal{L}_4,5,6)\\
&=-\braket{1256}\mathcal{I}^{\alpha\beta}\mathcal{I}_{\gamma\delta}\frac{\epsilon_{\beta}(\cdot,4,5,6)Z_6^{\delta}}{\braket{\mathcal{I}56}\braket{\mathcal{I}6(12)\cap \bar{5}}}\,,
\end{align}
where we used that $\mathcal{L}_4=(6,(12)\cap \bar{5})=(6,Z_4\braket{5612}+Z_5\braket{6412})$. We therefore observe that 
\begin{equation}
    \braket{1256}=0\,\,\implies \,\, (l_4)_{a\dot{a}}=0\,.
\end{equation}
We still have one degree of freedom left in our second cut solution which has $\mathcal{L}_3=(4,l_{56}(\beta))$. One can readily show that for codimension one solutions to the Landau equations our choice $\mathcal{L}_1=(1,2)$ requires $\alpha_{11}=0$. One way to see this is that all of the momenta in the first Kirchhoff equation are proportional to $\mathcal{I}_{\gamma\delta}Z_1^{\delta}$ except for $(l_1+p_2-l_2)\propto I_{\gamma\delta}Z_2^{\delta}$; hence either this latter momentum must be zero, which we do not have the remaining degrees of freedom available to achieve, or $\alpha_{11}=0$. Setting $\alpha_{11}=0$ requires that either all of the remaining $\alpha$'s in the third Kirchhoff equation are zero, or that the remaining vectors satisfy the determinant condition
\begin{gather}
    \text{Det}\big[l_2-p_2,l_2,l_2+p_3,l_1+p_2-l_2\big]\Big\vert_{\mathcal{L}_2=(2,4)}=0\\
    \implies \braket{1234}\braket{234l_{56}}=0\,.\label{eq:secvanish}
\end{gather}
Let us consider the case where not all of the remaining of $\alpha$'s in the third Kirchhoff equation are zero, in which case (\ref{eq:secvanish}) applies, and we can solve this requirement by choosing
\begin{equation}
    l_{56}=(56)\cap\bar{3}\,.
\end{equation}
Plugging this solution into the third Kirchhoff equation we find that this choice sets two of the momenta collinear,
\begin{equation}
    \mathcal{L}_2=(2,4)\And \mathcal{L}_3=(4,(56)\cap{3})\implies l_2+p_3\parallel l_2+p_3-l_3+p_4\label{coll1}\,.
\end{equation}
Plugging in our now fully localized cut solution into the fourth Kirchhoff equation we find that two of the momenta are automatically collinear,
\begin{equation}
    \mathcal{L}_2=(2,4)\And \mathcal{L}_3=(4,(56)\cap{3})\implies l_3-p_4\parallel l_2+p_3-l_3+p_4\label{coll2}\,,
\end{equation}
which is as expected from momentum conservation at the vertex where the momenta~(\ref{coll1}) and~(\ref{coll2}) meet. In summary we have found the solution
\begin{equation}
\begin{gathered}
       \mathcal{L}_1=(1,2),\quad \mathcal{L}_2=(2,4),\quad \mathcal{L}_3=(4,(56)\cap\bar{3}),\quad \mathcal{L}_4=(6,(12)\cap\bar{5})\\
    \text{with the constraint:   }\braket{1256}=0 
\end{gathered}\label{eq:examplesol}
\end{equation}
which solves the Kirchhoff equations,
\begin{equation}
    \begin{gathered}
   l_1=l_4=0\\
   \alpha_5 (l_2+p_3)+\alpha_{12}(l_2+p_3-l_3+p_4)=0,\quad \alpha_6(l_3-p_4)-\alpha_{12}(l_2+p_3-l_3+p_4)=0
    \end{gathered}\label{eq:examplesystem}
\end{equation}
by setting the momenta on the first line soft, and all three momenta on the second line collinear. Lastly, we must verify that our solution (\ref{eq:examplesol}) is not a subcase of a solution which should be associated to an IR-divergence. As explained in the previous sections, IR divergences are associated to solutions to the Landau equations which do not impose any constraints on the external kinematics. We note that it is impossible to solve $l_1=l_4=0$ on the second cut solution in Table~\ref{tab:firstonshellsol} without imposing $\braket{1256}=0$, thus our solution (\ref{eq:examplesystem}) cannot be associated to an IR divergence. This concludes our prototypical example of a leading solution to the Landau equations.

\subsection{Relaxations}
\label{sec:relaxations}

In order to complete the enumeration of singularities of ${\cal{G}}_7$, it remains to consider all possible relaxations of graph Fig.~(\ref{fig:two})(b). By relaxation, we mean any subgraph that can be obtained by contracting any subset of propagators (which, at the level of the Landau equations, amounts to setting the corresponding Feynman parameters to zero). We don't need to explicitly consider ``edge deletions'', mentioned at the end of Sec.~\ref{sec:zig}, since solutions associated to such subgraphs come along for the ride as solutions of the parent graph with identically zero momentum flowing through the deleted edge.

It is interesting to note that the analysis of the four-, five- and six-particle graphs reviewed in Sec.~\ref{sec:two} does not reveal any additional singularities in relaxations beyond those already encountered at leading order. This certainly is not the case for ${\cal{G}}_7$ since it does not have any leading singularities corresponding to symbol letters of type $b_6$, yet it contains (after relaxing 10 propagators) subgraphs of one-loop three-mass box type, which do have such singularities (see for example~\cite{Dennen:2015bet}). Therefore we know that ${\cal{G}}_7$ has singularities (leading and/or subleading) corresponding to \emph{all} elements of the heptagon symbol alphabet, but our analysis is not yet enough to conclude that no \emph{other} singularities are present. Since ${\cal{G}}_7$ has quite a few nontrivial graphs as relaxations, we postpone a full analysis of this question to a companion paper.

\section{Conclusion}

Landau singularities of Feynman integrals can become complicated very quickly as the number of loops and external points increases. Nevertheless, in this paper we have been able to determine analytically (guided by a bit of numerical experimentation) the (first-type, leading) Landau singularities of the Feynman integral associated to the planar, massless, seven-point graph ${\cal{G}}_7$ shown in Fig.~\ref{fig:two}(b). We have found that its singularity locus is surprisingly simple, and actually corresponds to (a subset of that of) the heptagon symbol alphabet of~\cite{Drummond:2014ffa}. Our work was motivated in part to begin an exploration of the kinds of singularities that can occur in seemingly complicated massless planar seven-point graphs such as ${\cal{G}}_7$, and to provide thereby some independent verification of one of the main assumptions underlying the bootstrap of seven-point amplitudes in $\mathcal{N}=4$ super-Yang-Mills theory. Although we now know that ${\cal{G}}_7$ has subleading singularities outside the heptagon alphabet~\cite{Lippstreu:2023oio}, it remains an interesting open problem to gather more data about the kinds of singularities that can appear in seven-point amplitudes and integrals, and especially to see if this set admits any universal characterization that could aid the bootstrap program in SYM or more general theories.

\acknowledgments

We are grateful to I.~Prlina and S.~Stanojevic for collaboration in the early stages of this work, to S.~Mizera for correspondence, and to A.~Ghandikota for independently checking the graphical reduction shown in Fig.~\ref{fig:three} as part of his summer 2022 SPRINT{$|$}UTRA project. This work was supported in part by the US Department of Energy under contract DE-SC0010010 (Task F) and by Simons Investigator Award \#376208.

\end{document}